\documentclass[pre,notitlepage,nofootinbib,twocolumn,superscriptaddress]{revtex4-1}
\usepackage[utf8x]{inputenc}
\usepackage{amsmath}
\usepackage{amssymb}
\usepackage{graphicx}
\usepackage{enumitem}
\usepackage{todonotes}
\usepackage[caption=false]{subfig}

\newcommand{\ie}{{\it i.e. }}

\newcommand{\I}{\mathbb{I}}
\newcommand{\tr}{\text{tr}}

\usepackage{bm}
\renewcommand{\vec}[1]{\boldsymbol{\mathbf{#1}}}
\usepackage[colorlinks=true,citecolor=blue,linkcolor=magenta]{hyperref}

\graphicspath{{./}{./images/}}

\begin{document}

\title{Operator dynamics in Brownian quantum circuit}

\author{Tianci Zhou}
\email{tzhou@kitp.ucsb.edu}
\affiliation{University of Illinois, Department of Physics, 1110 W. Green St. Urbana, IL 61801 USA}
\affiliation{Kavli Institute for Theoretical Physics, University of California at Santa Barbara, CA 93106, USA}

\author{Xiao Chen}
\email{xchen@kitp.ucsb.edu} 
\affiliation{Kavli Institute for Theoretical Physics, University of California at Santa Barbara, CA 93106, USA}

\date{\today}
\begin{abstract}
  We view the operator spreading in chaotic evolution as a stochastic process of height growth. The height of an operator represents { the size of its support} and chaotic evolution increases the height. We consider $N$-spin models with all 2-body interactions and embody the height picture in a random model. The exact solution shows that the mean height, being proportional to the squared commutator, grows exponentially within $\log N$ scrambling time and saturates in a manner of logistic function. We propose that the temperature dependence of the chaos bound could be due to initial height biased towards high operators, which has smaller Lyapunov exponent. 
\end{abstract}

\maketitle

\section{Introduction}

Quantum many-body chaos recently has drawn attention from many fields, including quantum gravity\cite{Shenker2013a}, quantum information\cite{hosur_chaos_2016} and condensed matter physics\cite{blake_thermal_2017}. The dynamics of chaos can be diagnosed by the squared commutator (out-of-time-order correlator)\cite{Larkin1969}
\begin{equation}
  \label{eq:c_of_t}
C(t)=-\langle [V(t),W]^2\rangle_{\beta} = -  \tr(  [ \rho^{\frac{1}{4}} V(t) \rho^{\frac{1}{4}},W]^2 )
\end{equation}
where $V(t) = e^{iHt}V(0)e^{-iHt}$ and $\rho = \frac{e^{- \beta H}}{\tr( e^{- \beta H} ) }$. In some strongly chaotic systems, $C(t)$ can increase exponentially in time as $e^{\lambda t}$. The positive number $\lambda$ is the quantum analogy of the Lyapunov exponent\cite{Sekino2008,Shenker2013a, Shenker2015, Kitaev2015, Maldacena2015, Maldacena2016}, which measures ``exponential divergence of trajectories''. Nevertheless there is a quantum upper bound $\frac{2\pi}{\beta}$ for $\lambda$\cite{Maldacena2015} ({ only for the thermally regulated version in Eq.~\eqref{eq:c_of_t}, not the thermal average of the squared commutator\cite{liao_nonlinear_2018}}), which is known to be saturated by fast scramblers such as black hole and the Sachdev-Ye-Kitaev (SYK) model\cite{Sachdev1993, Kitaev2015, Maldacena2015, Maldacena2016}. 

In Eq.~\eqref{eq:c_of_t}, one can view the $W$ as probing the content of the Heisenberg evolved operator $V(t)$. The failure for it to commute with $V(t)$ indicates that the (spatial) support of $V(t)$ has spread to $W$. Recently there are many effective hydrodynamic models about operator spreading\cite{Lucas2017,Blake2018,roberts_operator_2018,keyserlingk_operator_2017,nahum_operator_2017}. Among these works, the study of random unitary circuits\cite{keyserlingk_operator_2017,nahum_operator_2017} provides a biased random walk picture that successfully captures the growth of $C(t)$ in systems with {\rm local} interactions. %

\begin{figure}[h]
\centering
\includegraphics[width=0.8\columnwidth]{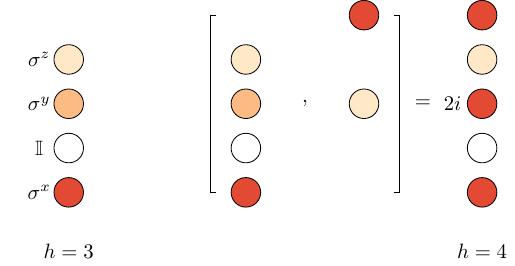}
\caption{Height of a Pauli string. (Left) A Pauli string consisting of tensor products of Pauli matrices. The height is the number of Pauli matrices. (Right) The commutator with an overlapping interaction term can increase the height by $1$. }
\label{fig:ball}
\end{figure}

Inspired by these works, we formulate the operator spreading as a classical height growth process. We define the height in a simple example in Fig.~\ref{fig:ball}, which for Pauli string (tensor product of Pauli matrices and identity on different sites) is the number of Pauli matrices.  %
The time evolution in a short interval can be approximated by the commutator, which for 2-body interactions can change the height by one. An operator generally is a superposition of Pauli strings. It then hosts a probability distribution of height. Its time evolution is a transition from one height to the other, whose rates are fixed by the number of available terms participating the commutator in Fig.~\ref{fig:ball}. In order for the operator to grow, { the two-body interaction must have one part inside the support of the operator and another outside}. This is very similar to the facilitated dynamics in the kinetically constrained model\cite{ritort_glassy_2003}, which is used to describe the classical glass dynamics. 

In this paper,  we consider a quantum dot with all to all 2-body interactions -- a zero dimensional model. Generalization of the height picture to higher dimensions will be presented elsewhere. We use a toy model called Brownian quantum circuit to analytically derive and solve the master equation for the height transition described above. We find that $C(t)$ is the mean height. Starting from a simple (single site) operator, the dynamics has three stages: (1) the initial exponential growth of the mean height before scrambling time of order $\log N$, where $N$ is the number of sites, (2) the slow down of the operator growth caused by finite $N$ and large height fluctuation in the intermediate time and (3) the exponential decay to the equilibrium height with the decay rate the same as the initial Lyapunov exponent.

We believe the height picture as demonstrated by the brownian quantum circuit calculation is generic for non-integrable systems { with $k$-local interactions} at infinite temperature. { With small modification, the height picture can interpret the physics of operator spreading with local and long range interactions in 1d and higher dimensions.}
The brownian circuit technique allows us to use randomness as tool to overcome the difficulty of the non-integrability. The idea dates back to the invention of random matrix theory, and recently resurfaced for instance in works about Sachdev-Ye-Kitaev model\cite{Sachdev1993, Kitaev2015, Maldacena2015, Maldacena2016} and random unitary circuits\cite{keyserlingk_operator_2017,nahum_operator_2017}.

We further find that the exponential growth rate is smaller for higher initial operators. This mechanism is possibly related to the temperature dependence of $\lambda_L$, where low temperature corresponds to higher initial operators, suppressing $\lambda_L$. After the submission of the paper, we are aware of an explicit demonstration of this mechanism in Ref.~\onlinecite{qi_quantum_2018}, which computes the height of operator at finite temperature in the large-$q$ SYK model.

\section{The Height Distribution and Operator Dynamics}
\label{sec:op_stat}
The height of a Pauli basis $B_j$ is the number of non-identity single site operators (Pauli matrices) in it. For a generic operator $V(t)$, we expand it in the Pauli basis $B_j$,
\begin{equation}
\label{eq:Ot-basis}
V(t) = \sum_j \alpha_j(t) B_j.
\end{equation}
Here we treat the operator space as a Hilbert space with inner product $\langle O_1, O_2  \rangle  = \frac{\tr( O_1^\dag O_2 )}{\tr(\I)}$. Due to unitarity, $|\alpha_j(t)|^2$ can be interpreted as the probability of the basis $B_j$, whose sum is normalized to $1$. The height distribution of $O(t)$ can thus be defined as
\begin{equation}
 f( h, t ) = \sum_{\text{height}(B_j) = h} |\alpha_j(t)|^2.
\end{equation}
We further define a $N+1$ component unit normalized vector $\vec{f}$, whose $k$th component $f_k$ is $f(h =k, t )$. In the continuum limit $f(h, t)$ is the height probability density.
Chaotic evolution will quickly mix the operator to be evenly distributed in each height space. With this assumption (which can be dropped in the Brownian quantum circuit), we can show that
 ({$\mathcal{N} \sim \tr(\I)$, see App.~\ref{app:C_h_aver}})

\begin{equation}
C(t) \equiv  -\frac{1}{\mathcal{N}} \tr( [V(t), W]^2 ) = \frac{\langle h(t) \rangle }{N}, 
\end{equation}
namely the squared commutator is proportional to the mean height. Therefore the operator scrambling is encoded in the height distribution. 

In the following, we will use the general $q\times q$ Pauli matrices $\sigma^{\mu}$, which is the generators of ${\rm SU}( q^2 )$. The $q \rightarrow \infty$ is a helpful limit in checking the results.

 \section{Operator Dynamics of Brownian Quantum Circuit}
\label{sec:op_dyn}
\subsection{Brownian Quantum Circuit}

We aim to understand the operator dynamics for systems with generic all to all 2-body interactions. Randomness is our key tool to construct such generic non-integrable yet still solvable models.

We put spacetime randomness in the strength of the 2-body spin-spin interactions. The model is called the Brownian quantum circuit\cite{lashkari_towards_2013}. In a short interval of $\Delta t$, the circuit (or the time evolution) performs a random walk on the unitary group. The direction of the displacement is specified by the Hamiltonian, which is spanned by the 2-body interactions. More concretely, the evolution in a short interval is governed by the Hamiltonian
\begin{equation}
  H_s = J \sum_{i<j} \sum_{\mu_i, \mu_j =0}^{q^2-1}   \sigma_i^{\mu_i} \otimes  \sigma_j^{\mu_j} \Delta B^s_{i, j, \mu_i, \mu_j},
\end{equation}
where $\sigma_i^{\mu_i}$ are the generalized Pauli matrices on site $i$ and $\sigma_i^0 = \I$. The strength $B^s_{i,j,\mu_i,\mu_j}$ of the spin-spin interactions over sites $i$ and $j$ is a Gaussian random variable with variance proportional to $\Delta t$. The continuum limit of the evolution
\begin{equation}
e^{- i H_s \Delta t } e^{- i H_{s-1} \Delta t }   \cdots  
\end{equation}
defines the Brownian quantum circuit.

We use the normalization (different from Ref.~\onlinecite{lashkari_towards_2013}) $J = \sqrt{ \frac{2}{q^4 N}} $ where $N$ is the total number of sites. This makes the mean field experienced by each spin to be order $\mathcal{O}(1)$. 

 \subsection{Master Equation of the Height Distribution}

The Brownian quantum circuit does not distinguish between different types of Pauli matrices, hence on average it gives a clean and simple equation in the height space. The master equation of the height distribution $f_k( t )$ is given by 
\begin{equation}
\label{eq:prob-evo}
  \frac{d \vec{f}(t)}{d t} = A_f \vec{f}(t)
\end{equation}
with a tri-diagonal stochastic matrix $A_f$
\begin{equation}
\label{eq:A_f}
\begin{aligned}
(A_f)_{k,k} &= \frac{4}{N}  k \big[-(N-k) + \frac{1}{q^2} ( N - 2k + 1 ) \big] \\
(A_f)_{k-1,k}& =  \frac{4}{N}\frac{k(k-1)}{q^2} \\
(A_f)_{k+1,k} &= \frac{4}{N} k (N - k)\big[ 1 - \frac{ 1}{q^2}  \big] .\\
\end{aligned}
\end{equation}

The equation is derived by adopting and developing the It\^o calculus techniques from Ref.~\onlinecite{lashkari_towards_2013}, see details in App.~\ref{app:deriv_master}. It has recently been derived as a special case of the Brownian cluster model in Ref.~\onlinecite{xu_locality_2018} by a different approach.

The evolution matrix $A_f$ has zero column sum and preserves the total probability. Its surprisingly simple tri-diagonal structure is a consequence of purely two-body interactions in the Hamiltonian: the interaction terms can only change the height by one in an infinitesimal step. A particularly simple limit is $q \rightarrow \infty$ that $A_f$ is a lower triangle matrix. Its off-diagonal element $(A_f)_{k+1,k}$ is $\frac{4}{N} k (N - k)$ which is proportional to the number of the height increasing commutators shown schematically in Fig.~\ref{fig:ball}. At finite $q$, the number of interaction terms that commutate with the operator is proportional to $\frac{1}{q^2}$. These terms are responsible for the process of decreasing the height by one, which is consistent with the $\frac{1}{q^2}$ dependence in the upper diagonal element of $A_f$. This analysis is essentially the same as the case of a random unitary gate\cite{nahum_operator_2017,keyserlingk_operator_2017}.

The height probability has two linearly independent stationary solutions. The first one is the identity operator
\begin{equation}
\begin{aligned}
f_0 &= 1 \quad f_k = 0 \quad k \ge 1 \\
\end{aligned}
\end{equation}
as the identity is invariant under unitary evolution. In fact the probability of identity and non-identity operators are separately conserved. The non-identity sector containing those operators of none-zero height will be driven to
\begin{equation}
f_0 = 0 \quad f_k = \frac{{N \choose k } (q^2 - 1)^k }{q^{2N} - 1}  \quad k \ge 1  
\end{equation}
where $f_{k\ne 0}$ is the ratio of the height $k$ operators with respect to the all in this sector. Therefore, the ultimate fate of the non-identity sector is an equal weight superposition of all non-identity operators, \ie a maximally random operator. The average height saturates to
\begin{equation}
\label{eq:h_sat}
h_{\rm sat} = \langle h \rangle = \frac{N(q^2 - 1)q^{2N-2}}{q^{2N} - 1} \simeq N ( 1 - \frac{1}{q^2} ) . 
\end{equation}

 \subsection{General Solutions of the Master Equation}
We solve the master equation analytically for any initial conditions. To express the result, we define
\begin{equation}
\lambda_q = 4 ( 1- \frac{1}{q^2}),
\end{equation}
where $\lambda_q$ is the Lyapunov exponent (see below). %

At early time, or large $N$ limit at fixed time, the master equation simplifies to
\begin{equation}
  \frac{d f_k(t)}{d t} = - \lambda_q k f_k + \lambda_q (k-1) f_{k-1}.
\end{equation}
The coefficients of $f_{k-1}$ is the rate of height increase, which is proportional to the height itself, indicating an initial exponential growth. Similar equation\cite{roberts_operator_2018} was proposed for the height growth in the Sachdev-Ye-Kitaev (SYK) model\cite{Sachdev1993, Kitaev2015}. Here we write the solution in terms of its generating function
\begin{equation}
\label{eq:f_gen_early_t}
\sum_{k=0}^{\infty} f_k(t) z^k = \sum_{k=0}^{\infty} \left( \frac{ze^{-\lambda_q t}}{1  -z ( 1- e^{-\lambda_q t} ) } \right)^{k} f_{k}( t =0 ).
\end{equation} 
All the moments can thus be computed. However, this only works for early time with distributions localized at small height, \ie $\langle h \rangle \ll N$. 

To understand the physics beyond early time, we take the continuum limit of the master equation and get
\begin{equation}
\label{eq:cont-master}
\partial_t f( h, t ) = - \frac{4}{N} \partial_h[ h ( h_{\rm sat} - h ) f ].
\end{equation}
The flux of the probability density vanishes at $h = 0$ and $h = h_{\rm sat}$, conserving the total probability\footnote{Strictly speaking, this discrete equation allows to generate operators higher than $h_{\rm sat}$. However, those exceptions are exponentially small (even for small $N$) as can be seen from the steady state distribution.}. Please change the sentence to "After multiplying Eq. 15 by h on both sides and integrating, we see that
the mean height obeys ... ".

After multiplying Eq.~\eqref{eq:cont-master} by $h$ on both sides and integrating, we see that the mean height obeys
\begin{equation}
\label{eq:logistic}
\begin{aligned}
\partial_t \langle h \rangle  = \frac{4 }{N} ( h_{\rm sat}  - \langle h \rangle )  \langle h \rangle  - \frac{4}{N}( \langle h^2 \rangle - \langle h \rangle^2   ). 
\end{aligned}
\end{equation}
If the fluctuation of the height is much smaller than $N$, then the mean height evolves according to the logistic equation\cite{chen_operator_2018}. This mean field picture again corresponds to Fig.~\ref{fig:ball} where the growth of height is proportional to the number of interaction terms increasing the height. The solution is
\begin{equation}
\langle h(t) \rangle  = y( \langle h(0) \rangle , t ),
\end{equation}
where $y(h,t)$ is the logistic function
\begin{equation}
y(h, t ) = \frac{ h_{\rm sat} h  e^{ \lambda_q t}}{h_{\rm sat}  +h  (e^{\lambda_q t}-1) }. 
\end{equation}
This mean field solution was heuristically argued in Ref.~\onlinecite{chen_operator_2018}, under the assumption of vanishing fluctuation and essentially $q = \infty$ limit. The fluctuation turns out to be important at intermediate time (see Fig.~\ref{fig:h_compare.pdf}). %

The general solution can be obtained by the method of characteristics. It is conveniently expressed as
\begin{equation}
\begin{aligned}
  f(h,t)= &f(y^{-1}(h, t=0 ), t=0 ) \frac{d y^{-1}(h,t)}{d h}, 
\end{aligned}
\end{equation}
whose moments are
\begin{equation}
\label{eq:h_aver_conti}
\langle h^l(t) \rangle  = \int_0^N f(h, t )  h^l dh  = \int_0^N f( h, 0) [y(h, t)]^l dh. 
\end{equation}

\section{Scrambling and Dynamics of the Height Distribution}
\label{sec:result}
\subsection{Evolving from a Simple Operator}
\label{subsec:simple_op}
We first study the height distribution evolved from a simple operator, which is localized at one site. It can be modeled by a delta distribution for the discrete master equation or an exponentially localized distribution $e^{-h}$ in the continuum limit. As the evolution goes on, the mean height increases to an appreciable fraction of the maximal height in a very short period of time and then slows down on the way to the steady state distribution. 

We obtain the exact solution\footnote{Exact in $N \rightarrow \infty$. Relative difference $< 0.1\%$ compared to the numerical data of $N = 10^4$} by Eq.~\eqref{eq:h_aver_conti}
\begin{equation}
\label{eq:h_in_Ei}
\langle h(t) \rangle = h_{\rm sat} (1 - s e^s {\rm Ei}[-s] ) ,
\end{equation}
where $\rm Ei[s]$ is the standard exponential integral function and $s$ is a parameter
\begin{equation}
s = h_{\rm sat} \exp( - \lambda_q t ) 
\end{equation}
that separates the evolution into three different time regimes.

\begin{figure}[h]
  \centering
  \includegraphics[width=0.8\columnwidth]{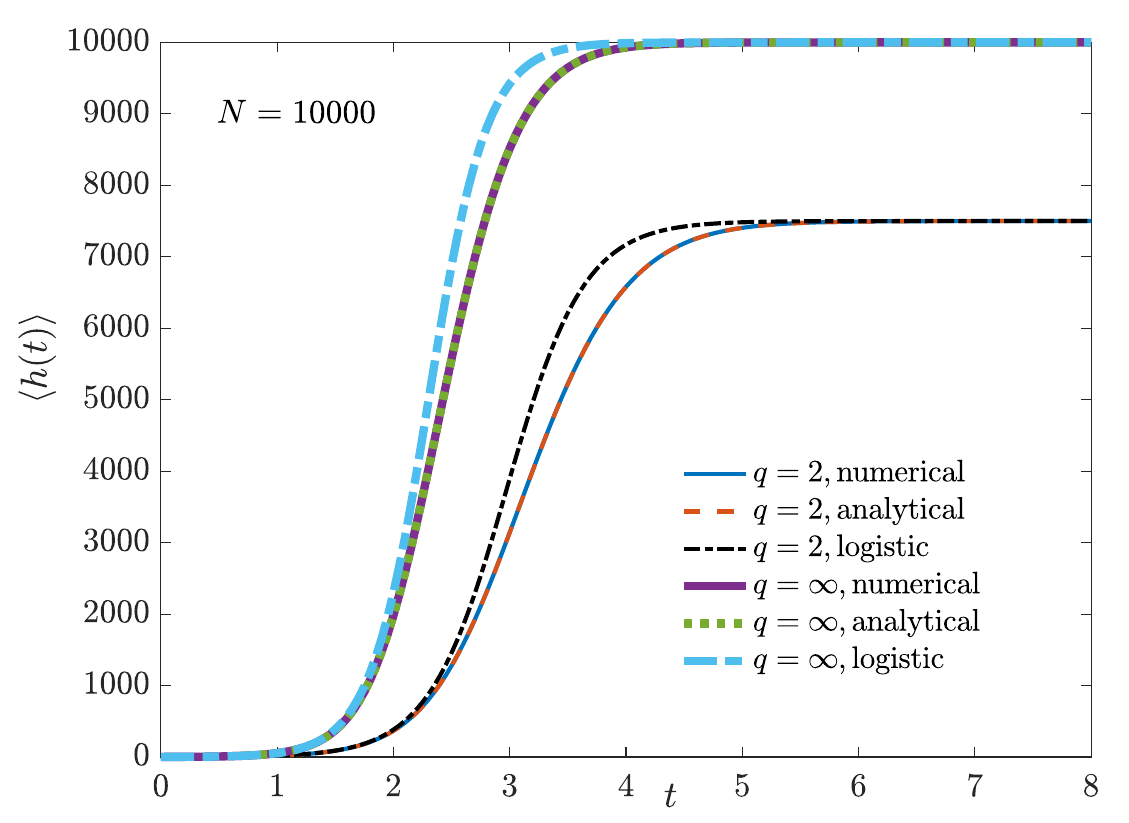}
  \caption{Comparison of the mean height with logistic function (mean field) and analytic solution in Eq.~\eqref{eq:h_in_Ei}. The large fluctuation in the intermediate time makes the dynamics slower than the logistic function. }
  \label{fig:h_compare.pdf}
\end{figure}

\begin{figure}
  \centering
  \subfloat[]{
    \label{fig:P_early.pdf}
    \includegraphics[width=0.8\columnwidth]{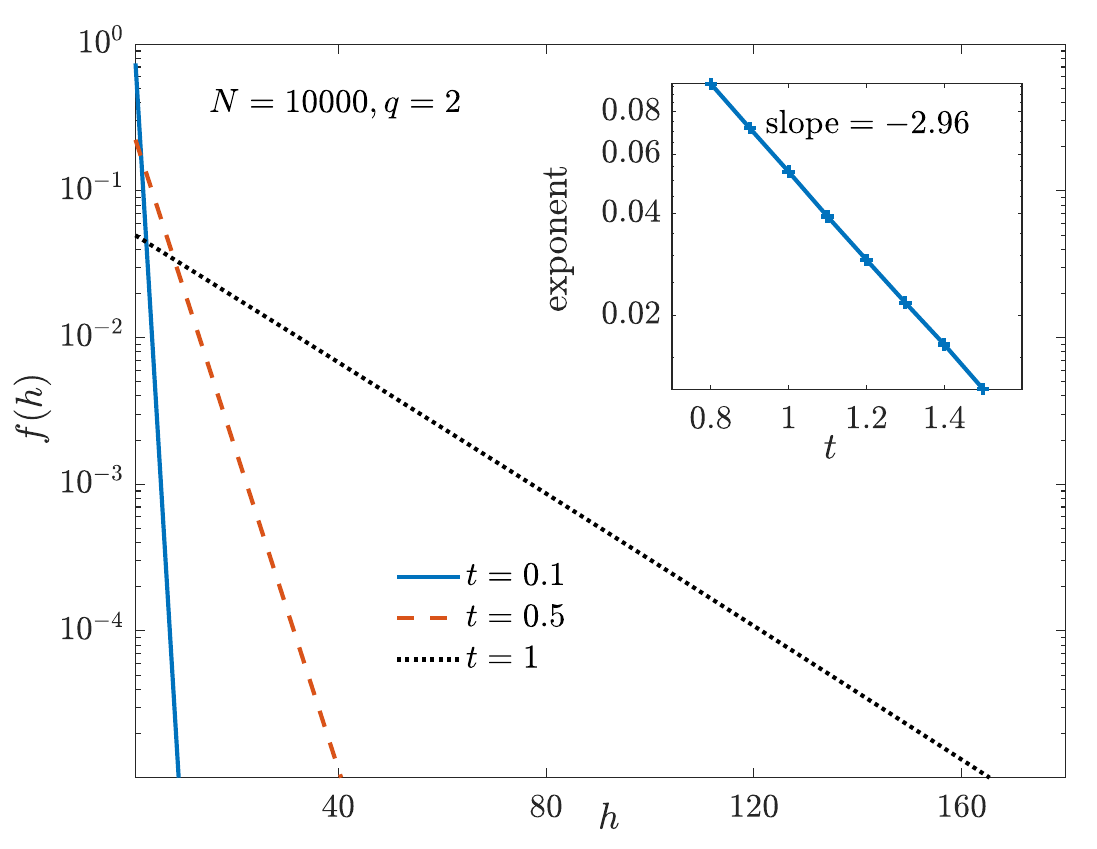}
  }\\
  \subfloat[]{
    \label{fig:P_late.pdf}
    \includegraphics[width=0.8\columnwidth]{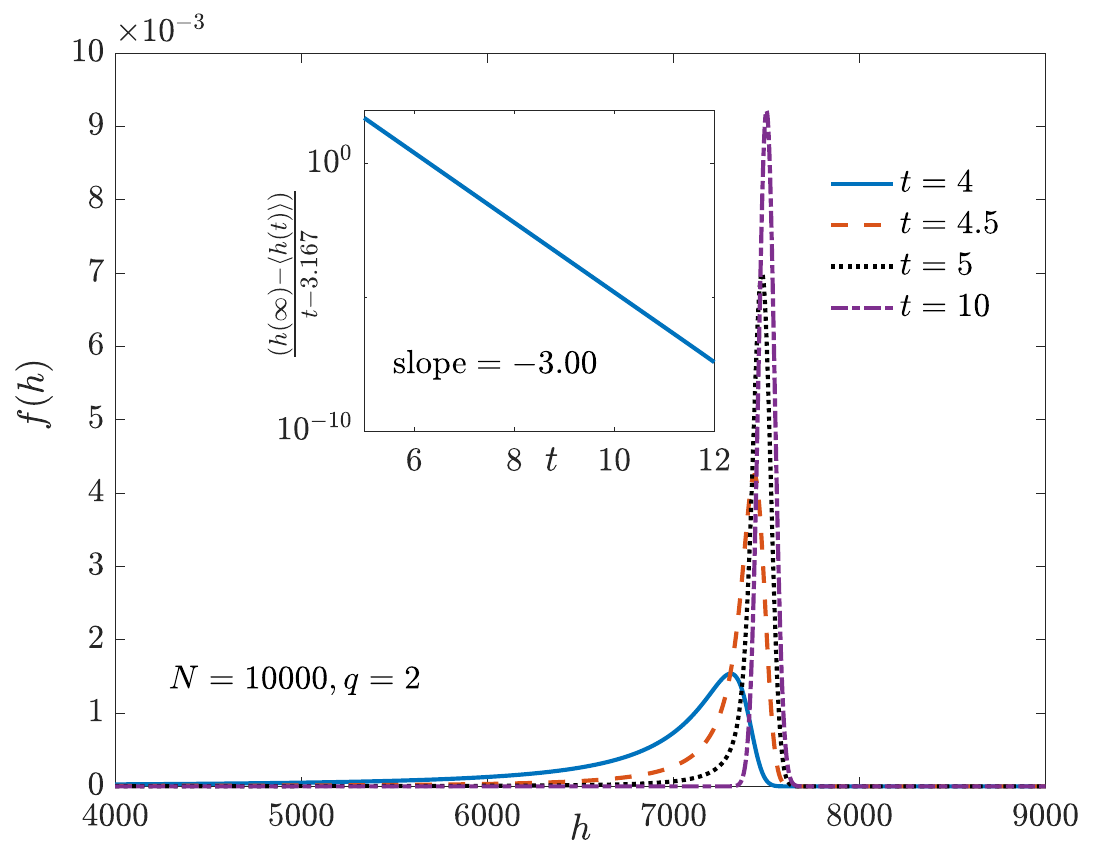}
  }\\
  \subfloat[]{
    \label{fig:P_mid.pdf}
    \includegraphics[width=0.8\columnwidth]{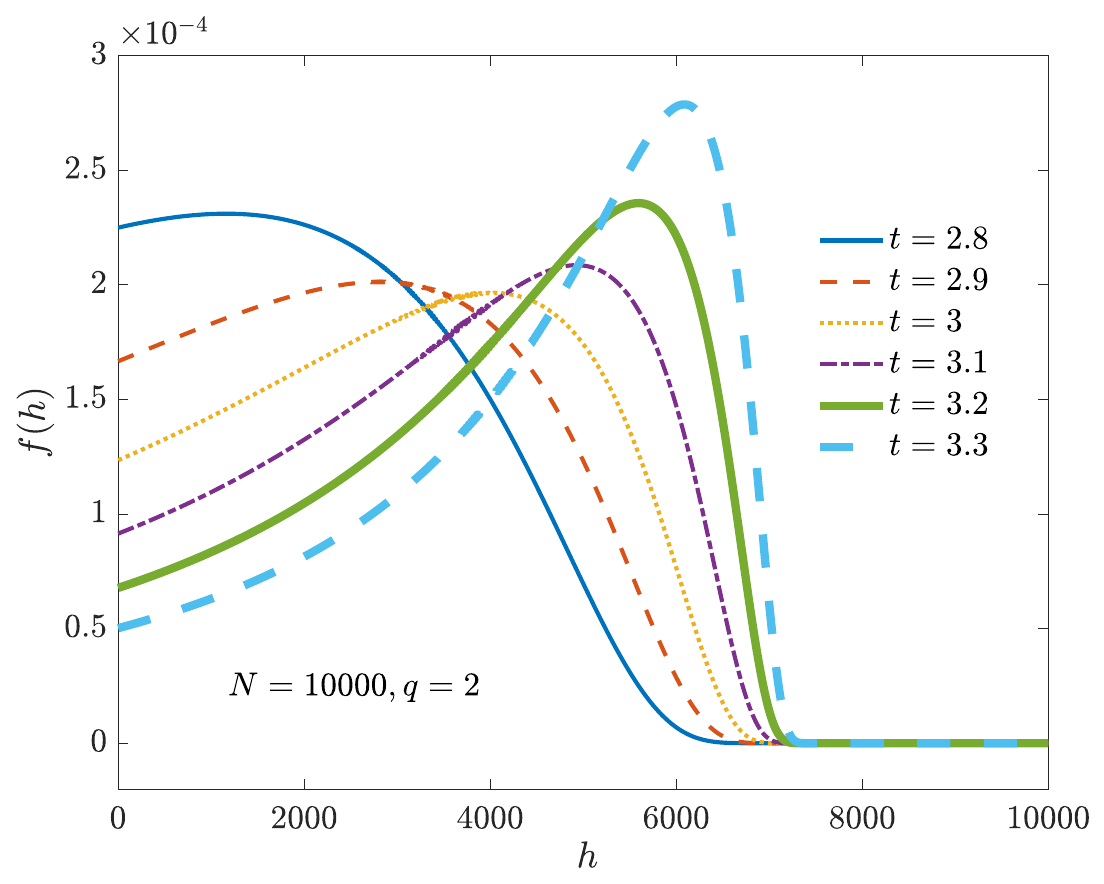}
  }
  \caption{The height distribution and mean height evolved from a simple operator localized at $h = 1$. (a) Early time: the profile looks like a collapsing sand pile. It is exponentially decreasing in space and its exponent is exponentially decreasing in time. Inset: numerically measured exponent is linearly proportional to $t$ on the semi-log scale with the slope $ -2.96$, close to $-\lambda_{q=2} = 3$. (b) Late time: the height distribution is exponentially decaying to the steady state of a random operator. Inset: The numerical verification of Eq.~\eqref{eq:C_t_exp}. (c) Intermediate time: the height distribution has appreciable value over almost the whole range of height. The height fluctuation is of order $N$. } 
  \label{fig:mean and std of nets}
\end{figure}

{\bf Early time:} $s \sim h_{\rm sat}$. This is the regime where the average height has not felt the existence of the saturation height (or $N$). By using the discrete solution in Eq.~\eqref{eq:f_gen_early_t}, the initial condition $f_k(t = 0 ) = \delta_{k1}$ gives
\begin{equation}
  f_k(t) = e^{ -\lambda_q t} [1 - e^{-\lambda_q t}]^{(k -1)}.
\end{equation}
The distribution profile looks like a collapsing sandpile whose surface (the probability) is exponentially decreasing in space. As time goes on, the surface rapidly becomes flatter as the exponent $\log ( 1- e^{-\lambda_q t}) \simeq e^{-\lambda_q t}$ also exponentially decays with time. These have been numerically confirmed in Fig.~\ref{fig:P_early.pdf}. 

The average height is growing exponentially with the same exponent
\begin{equation}
\label{eq:h_early_t}
\langle h(t) \rangle  = \partial_z \sum_{k=0}^{\infty}  z^k f_k(t)\Big|_{z = 1} = e^{\lambda_q t} \langle h(t=0) \rangle 
\end{equation}
{\it regardless of the initial condition} (as long as $\langle h \rangle / N \sim 0$). Hence the squared commutator
\begin{equation}
  C(t) = \frac{e^{\lambda_q t} }{N} \langle h(t= 0 ) \rangle 
\end{equation}
has exactly the same behaviors of many chaotic large-$N$ models \cite{Sachdev1993, Shenker2013a, Shenker2015, Kitaev2015, Maldacena2015, Maldacena2016}. This is an alternative way to show the scrambling time $\sim\log N$ and justifies to call $\lambda_q$ the Lyapunov exponent. The scaling can also be obtained by the large $s$ expansion of Eq.~\eqref{eq:h_in_Ei}, which gives the same $\langle h \rangle  \sim h_{\rm sat} \frac{1}{s} = e^{ \lambda_q t} $. 

{\bf Late time:} $s \ll 1$. This is the regime when the distribution is close to the steady state defined by a random operator. Fig.~\ref{fig:P_late.pdf} shows this approaching process. The small $s$ expansion of Eq.~\eqref{eq:h_in_Ei} gives
\begin{equation}
\label{eq:C_t_exp}
\begin{aligned}
&\langle h(t \gg \log N) \rangle \simeq h_{\rm sat} ( 1 + s \ln s + \gamma s )  \\
 &= h_{\rm sat} -  h^2_{\rm sat} (\lambda_q t-  \ln h_{\rm sat} -\gamma)  e^{- \lambda_q t} .
\end{aligned}
\end{equation}
It will saturate to $h_{\rm sat}$ exponentially with the decay rate the same as the initial growth. Notice that there is an extra linear $t$ correction term in front of $e^{-\lambda_qt}$, which is verified numerically in the inset of Fig.~\ref{fig:P_late.pdf}. 

{\bf Intermediate time:} $s \sim \mathcal{O}(1)$. The is the regime between the early time and late time. $f(h)$ is a broad distribution with a large fluctuation of order $\mathcal{O}(N)$ (see Fig.~\ref{fig:P_mid.pdf}). Because of this, the operator growth is slower than the logistic function (Fig.~\ref{fig:h_compare.pdf}). 

 \subsection{Power Law Initial Distribution and Thermal Operator}
\label{subsec:power_op}
It would be interesting to understand the bound of the Lyapunov exponent at finite temperature in this context, \ie the operator dynamics of $e^{-\beta H/4}V(t)e^{-\beta H/4}$. There is no notion of finite temperature in a Brownian quantum circuit because the energy is not conserved. Here we interpret the finite temperature effect as a change of the initial height distribution different from the previously discussed simple operators. 

\begin{figure}[h]
  \subfloat[]{
    \label{fig:h_vs_t_delta_h_0.pdf}
    \includegraphics[width=0.8\columnwidth]{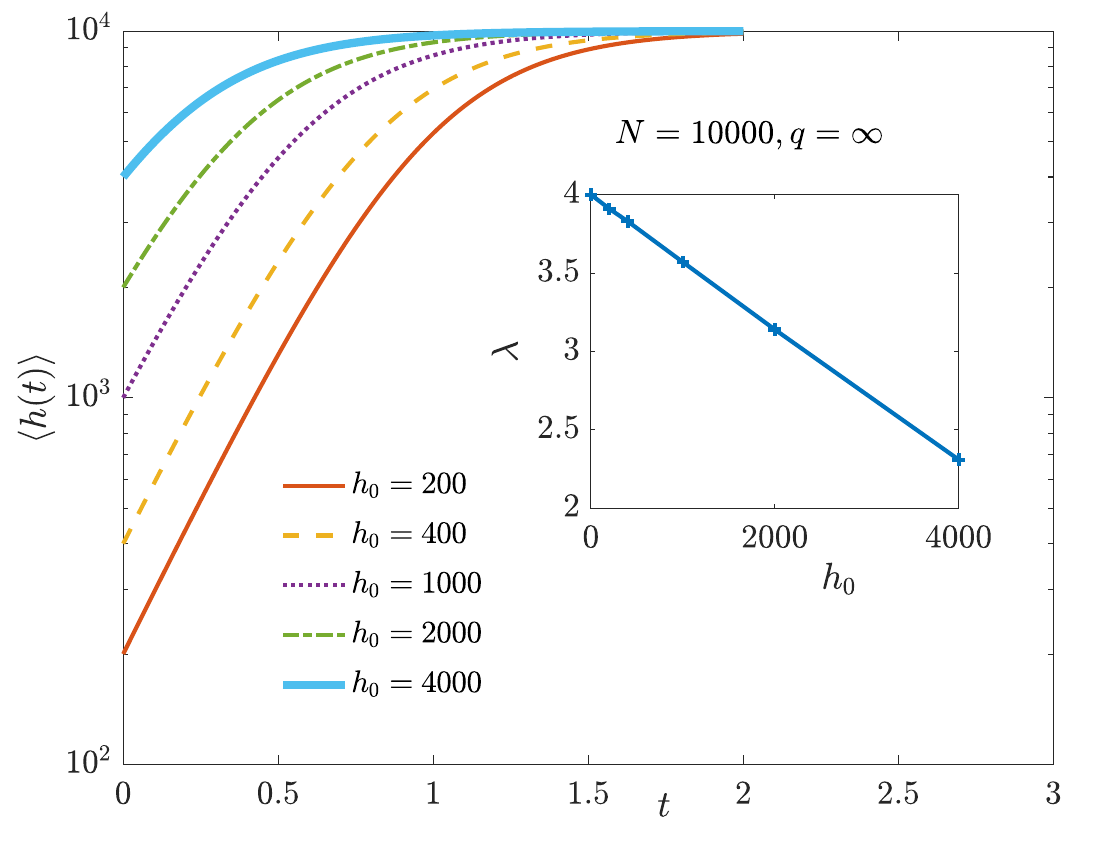}
  }\\
  \subfloat[]{
    \label{fig:h_vs_t_power.pdf}
    \includegraphics[width=0.8\columnwidth]{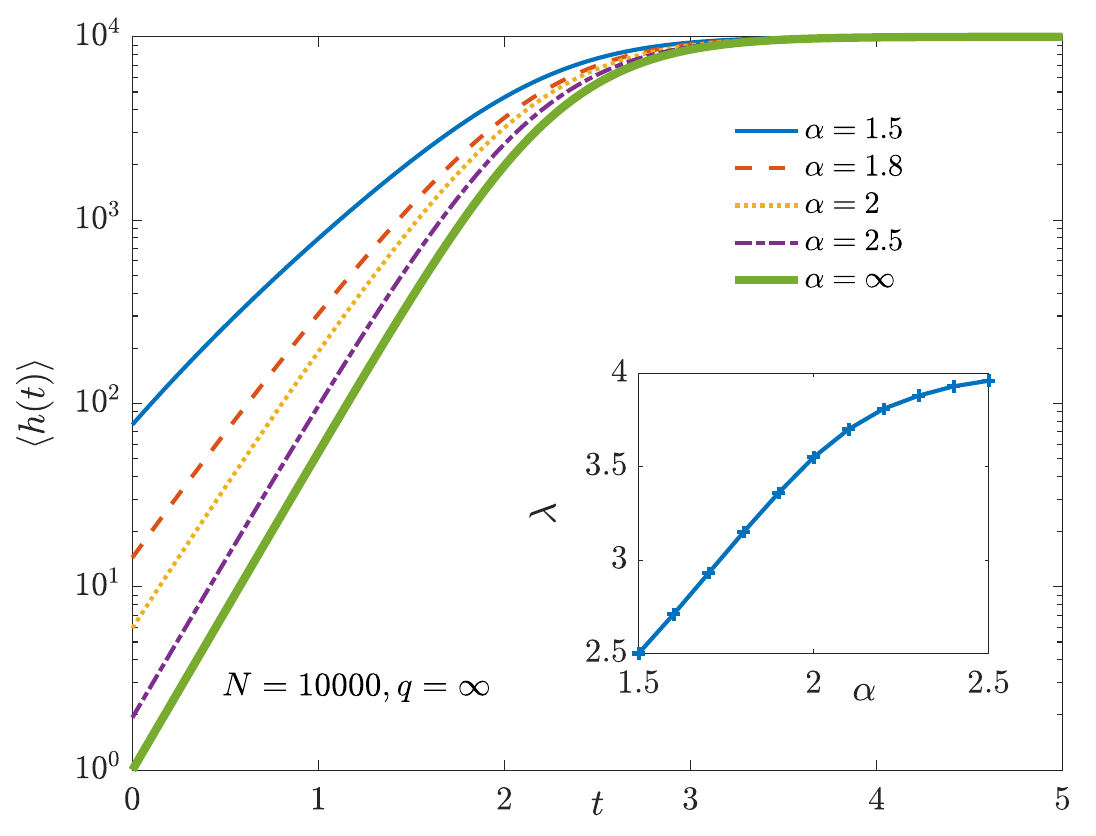}
  }
  \caption{Suppression of the growth rate for operator with initial $\langle h\rangle\gg 1$. (a) Growth exponent for operators initially localized at $h_0=a N$ with $a>0$. Inset: linear dependence of the exponent w.r.t $a$, saturating the bound around $t = 0$. (b) Growth exponent for power law decreasing initial condition $f_h \propto \frac{1}{h^\alpha}$. $\alpha = \infty$ is the exponential decreasing initial condition. Inset: growth exponent linearly depends on $\alpha$ and saturates to $\lambda_q$ when $\alpha > 2$. }
  \label{fig:slow_rate}
\end{figure}

As shown %
previously, 
the exponential growth rate in early time is always $\lambda_q$ (see Eq.~\eqref{eq:h_early_t}) when $\langle h \rangle\ll N $. The growth rate starts to become smaller when the ratio $\langle h \rangle/N$  is finite, suggesting a suppression of operator spreading in the Hilbert space. For example, when the initial distribution is $e^{ -(h - h_0)}\theta( h - h_0) $  where $h_0 \sim a h_{\rm sat} $ with $0 < a < 1$, we have
\begin{equation}
  \langle h(t) \rangle  = h_{\rm sat} [ 1  + s e^{ s+ h_0} {\rm Ei}( - s- h_0  ) ]. 
\end{equation}
The small $t$ behavior (large $s$ expansion) is
\begin{equation}
\label{eq:h_bound}
\langle h(t) \rangle \simeq \frac{h_{\rm sat} }{s+ h_0} \le \exp( \lambda_q( 1 - a)t ) .
\end{equation}
The exponent is therefore bounded by $\lambda_q ( 1- a)$. We numerically test the initial condition $f_k = \delta_{k h_0}$ in Eq.~\eqref{eq:prob-evo} and present the result in Fig.~\ref{fig:h_vs_t_delta_h_0.pdf}. We find that the exponential growth rate is smaller and saturates the upper bound in Eq.~\eqref{eq:h_bound} around $t=0$. 

This reveals a possible interpretation for the chaos bound at finite temperature -- the thermal operator is no longer localized at small height any more and should have more contribution from large height. At low temperature $\beta\gg 1$, if the Lyapunov exponent scales as $1/\beta$, according to the bound in Eq.~\eqref{eq:h_bound}, we speculate that the thermal operator has a large weight around the height $\sim h_{\rm sat}(1-1/\beta)$.

We further investigate the evolution from an initial power-law decaying distribution $f(h, t=0 ) \sim \frac{1}{h^{\alpha}}$, which favors longer operators than the localized distributions above. Its relation to the finite temperature physics is not justified, but this can always be viewed as a case study of possible slower operator growth. We test the range of $\alpha \in [ 1.5, 2.5]$  and fit the early time growth with
\begin{equation}
\langle h(t) \rangle  \simeq   \exp( \lambda( \alpha ) t ) .
\end{equation}
The inset of Fig.~\ref{fig:h_vs_t_power.pdf} shows that when $\alpha \in [ 1.5, 2]$, $\lambda$ is linearly proportional to $\alpha$ and approaches $\lambda_q$ when $\alpha > 2$.

\section{Conclusion}
\label{sec:conclusion}
In this work, we regard operator spreading in generic chaotic systems as a height growth process in Fig.~\ref{fig:ball}. The 2-local Brownian quantum circuit model gives an exact embodiment of the mechanism. We derive and solve the master equation governing the height transition of the operators for the full range of time. The squared commutator is the mean height in this formulation. 

We find that an initially simple operator (with $h=1$) will have an exponential growth with Lyapunov exponent $\lambda = 4 ( 1- \frac{1}{q^2})$, where $q$ is the Hilbert dimension of the spin. This growth will slow down around the scrambling time when the operator has a broad distribution over a large fraction in the operator space. It finally approaches the saturation value in a manner similar to the logistic function. %
While in the intermediate stage, the large fluctuation impedes the growth so that the dynamics is slower than the logistic function. 

We further give an interpretation of the temperature dependence of the Lyapunov exponent in the language of operator scrambling. We find that higher operators generically have smaller growth rate than shorter operators. Hence the chaos bound may come from an initial height distribution biased towards the high operators. We test the power law decreasing initial distribution $1/ h^{\alpha}$ and find a smaller growth rate with smaller $\alpha$. This however does not saturate the bound.

The height distribution picture and the associated Brownian quantum technique can be easily generalized to other types of interactions and dimensions. For example, there are recent works on the coupled cluster lattice models\cite{xu_locality_2018} in one dimensions and discussion about the bufferfly effect with the power-law decaying interactions\cite{chen_quantum_2018}. The similarity of the master equation with the Fredrickson-Andersen model\cite{fredrickson_kinetic_1984,fredrickson_facilitated_1985} relates the operator dynamics to the kinetically constrained model in glass physics, which is a new angle that can be explored in future works.

\acknowledgements  We acknowledge Leon Balents, Andreas W.W. Ludwig and Michael Stone for useful discussions. XS and TZ are supported by a postdoctoral fellowship from the Gordon and Betty Moore Foundation, under the EPiQS initiative, Grant GBMF4304, at the Kavli Institute for Theoretical Physics.
We acknowledge support from the Center for Scientific Computing from the CNSI, MRL: an NSF MRSEC (DMR-1720256). 

\appendix
\section{Relation of Squared Commutator to the Mean Height}
\label{app:C_h_aver}

We start from
the definition of $C(t)$ in the main text
and expand $V(t)$ in Hermitian basis $B_j$ (thus real $\alpha(t)$) %
\begin{equation}
V(t) = \sum_j \alpha_j(t) B_j.
\end{equation}
Only the basis non-commutative with $W$ will contribute. In general
\begin{equation}
  C(t) = - \frac{1}{\mathcal{N}} \sum_{h} \sum_{j,k} \alpha_j (t) \alpha_k (t) \tr( [B_j, W] [B_k, W] ), 
\end{equation}
For clarity, we specialize to the spin-$\frac{1}{2}$ model, taking $B_j$ to be product of Pauli matrices and $W = \sigma^x$ at site $0$. Then the cross term vanishes, and we organize the sum in order of height
\begin{equation}
\begin{aligned}
  C(t) &= - \frac{1}{\mathcal{N}} \sum_{h} \sum_{{\rm height}(B_j) = h} |\alpha_j(t)|^2 \tr( [B_j, W]^2 ) .
\end{aligned}
\end{equation}
We notice that the commutator is an anti-hermitian operator and minus sign makes $C(t)$ positive. As we assume in the text, $|\alpha_j(t)|^2$ for each basis $B_j$ with height $h$ are equal. This is by definition true for the Brownian quantum circuit model. Under this assumption
\begin{equation}
|\alpha^2_j(t)|\Big|_{ {\rm height}(B_j) = h} = \frac{f(h,t)}{3^h {N \choose h} } .
\end{equation}
Those basis non-commutative with $B_j$ must have $\sigma^{y,z}$ at site $0$, and there are $2 \times 3^{h-1} { N-1 \choose  h-1}$ such basis, which contributes equally. So
\begin{equation}
\begin{aligned}
C(t) &= \frac{8 \tr(\I) }{3\mathcal{N}} \sum_h f(h,t) \frac{h}{N} = \frac{8 \tr(\I) }{3\mathcal{N}}  \frac{\langle h(t) \rangle }{N}
\end{aligned}
\end{equation}
where we have chosen the convenient normalization factor $\frac{8 \tr(\I) }{3\mathcal{N}} = 1$. 

The extension to the thermal squared commutator is
\begin{equation}
\label{eq:C_beta_t}
\begin{aligned}
  C(\beta, t) &= -\frac{1}{\mathcal{N}} \frac{1}{Z}\tr( [e^{-\frac{\beta H}{4}} O(t) e^{ - \frac{\beta H}{4}} , V]^2 )  \\
  & = -\frac{1}{\mathcal{N}} \frac{1}{Z} \tr( [e^{iH t} ( e^{-\frac{\beta H}{4}} O(0) e^{ - \frac{\beta H}{4}}) e^{-iHt}  , V]^2 ) 
\end{aligned}
\end{equation}
The only change is the replacement of the initial operator $O(0)$ to its thermally regulated version $e^{-\frac{\beta H}{4}} O(0)e^{- \frac{\beta H}{4}}$.

\section{Derivation of the Master Equation}
\label{app:deriv_master}

In this appendix, we develop techniques in Ref.~\onlinecite{lashkari_towards_2013} to derive the discrete master equation of the height distribution. The key observation is that purity equation in Ref.~\onlinecite{lashkari_towards_2013} applies not just to density matrix $\rho$, but to any operators. 

We take the evolved operator $O(t)$ and compute its partial trace and average ``purity''
\begin{equation}
\phi_A = \frac{1}{q^N}\tr[ \tr_{\bar{A}}^2 (O(t))]
\end{equation}
As we assumed in the text, the initial operator is equally distributed on the basis of each height, such that the ``purity'' $\phi_A$ will only depend on the number of sites in region $A$. All the It\^o calculus computation in Ref.~\onlinecite{lashkari_towards_2013} follows and we get
\begin{equation}
 \frac{d \phi_k}{ dt} = \frac{k(N-k)}{N} \left\{  \frac{4}{q} \phi_{k-1} - 4\left(1 + \frac{1}{q^2} \right)\phi_k + \frac{4}{q} \phi_{k+1} \right\}
\end{equation}
where $\phi_k$ is the ``purity'' for arbitrary $k$ sites. 

We then introduce an intermediate variable: the cut averaged purity
\begin{equation}
\Phi_k( t) = q^{-(N-k)} \sum_{|A|=k} \phi_A(t) = q^{-(N-k)}{N \choose k} \phi_k(t)
\end{equation}
where the summation is over all the region containing $k$ sites. Through elementary counting, the height distribution is related to $\Phi$ as
\begin{equation}
{\boldsymbol \Phi} = {\rm PL}_{\tau}\,  \vec{f}
\end{equation}
where $({\rm PL}_\tau)_{ij} = { N-j \choose N-i}$ is the element (zero based index) of the rotated Pascal's lower triangle matrix. The master equation in matrix form is
\begin{equation}
 \frac{d {\boldsymbol f}}{d t} = A_{f} {\boldsymbol f}
\end{equation}
where $A_f =  {\rm PL}_{\tau}^{-1} A_{\Phi}  {\rm PL}_{\tau}$, the inverse matrix  ${\rm PL}_{\tau}^{-1}$ has element $(-1)^{i+j}{ N-j \choose N-i} $ and $A_{\phi}$ is tri-diagonal
\begin{equation}
\begin{aligned}
  (A_{\Phi})_{k,k} &= - \frac{4}{N} \left[1 + \frac{1}{q^2}\right]k ( N - k )  \quad &{\text{diagonal}} \\
  (A_{\Phi})_{k,{k-1}} &= \frac{4}{N}  (N - k ) (N-k +1 )  \quad &{\text{lower diagonal}} \\
  (A_{\Phi})_{k,{k+1}} &= \frac{4}{N} \frac{1}{q^2}  k ( k +1 )  \quad &{\text{upper diagonal}}.\\
\end{aligned}
\end{equation}

To derive the exact matrix element of $A_f$, we notice that the element of $A_{\Phi}$ can be obtained by taking derivatives, for example
\begin{equation}
k( N-k ) = \partial_x \partial_y x^k y^{N-k}\Big|_{x = y = 1} .
\end{equation}
We construct vector $A_{YX} = Z  {\rm PL}_{\tau}^{-1} YX  {\rm PL}_{\tau}$, where $X = {[x^0, x^1, \cdots, x^N]}$, $Y = [y^N, y^{N-1}, \cdots, y^0]^{\top}$, $Z = [z^0, z^1, \cdots, z^N ]$. By using the combinatorial property of the Pascal matrix, we have
\begin{equation}
(A_{YX})_k = \sum_{i=1}^N ( 1- z)^{N-i} z^i y^{N-i} (1+x)^{N-k} x^k . 
\end{equation}
Selecting out terms with the corresponding power (e.g. diagonal element corresponds to terms with total power of $x$ and $y$ to be $N$) and using the derivative trick, we can solve the matrix $A_f$. We find that $A_f$ is a tri-diagonal matrix with elements specified in %
Eq.~(8) in the main text.

\bibliographystyle{apsrev4-1}
\bibliography{chaos_scrambling,ref}
\end{document}